# Coherent Two-photon Backscattering and Induced Angular Quantum Correlations in Multiple-Scattered Two-Photon States of the Light

Nooshin M. Estakhri[1,2,3,4*], Theodore B. Norris[3,4]

[1]*Fowler School of Engineering, Chapman University, Orange, CA 92866, USA*

[2]*Institute for Quantum Studies, Chapman University, Orange, CA 92866, USA*

[3]*Department of Electrical Engineering and Computer Science, University of Michigan, Ann Arbor, Michigan 48109, USA*

[4]*The Gérard Mourou Center for Ultrafast Optical Science, University of Michigan, Ann Arbor, Michigan 48109, USA*

*We present the emergence of coherent two-photon backscattering, a manifestation of weak localization, in multiple scattering of maximally entangled pure and fully mixed two-photon states and examine the effect of entanglement and classical correlations. Quantum correlations in backscattering are investigated for finite three-dimensional disordered structures in the weak localization regime as well as systems of a small number of scatterers with specified spatial arrangements. No assumptions are made on the statistical behavior of the scattering matrix elements. Furthermore, we study the interplay between quantum correlations induced by multiple scattering and the correlations that may be present in the illumination fields, and how they are manifested in the output modes. We study the effect of the dimensionality of the entanglement and the angular distribution of the jointly measurable photon pairs on the emergence of enhancement and angular quantum correlations and show how quantum correlations can be used as a probe of the entanglement dimensionality. We show that by increasing the disordered material density, the width of the coherent two-photon backscattering cones increases, in accordance with the reduction of the mean free path length within the structure.*

## I. Introduction

The study of multiple scattering in disordered media is a long-established subject in the field of classical optics. Wave interference and coherent effects in these media result in important phenomena such as coherent backscattering (CBS) [1]-[5], Anderson localization [6], and the creation of speckle patterns [7]. CBS is a manifestation of the weak localization of waves in disordered media arising due to interference effects intrinsic to the wave nature of light. Physically, the constructive interference between time-reversed propagation paths leads to enhanced scattered intensity at the precise backscattering direction, referred to as the CBS cone. This coherent effect survives the ensemble-averaging over different realizations of the positions of the scattering centers, as the direct and time-revered paths always introduce the exact equal phase, and ideally results in a two-fold enhancement in intensity (i.e., ratio of two to one in the classical CBS cone). Away from the precise backscattering direction the averaged intensity gradually drops to the diffuse value.

In the last two decades, there has been a growing interest towards the interaction of quantum light (light exhibiting nonclassical correlations) with disordered structures in both theory and experiment. In an early theoretical work, Beenakker [8] studied the subject of thermal radiation and amplified spontaneous emission from a disordered absorbing slab and from a chaotic cavity using random-matrix theory. Subsequently, photon correlations and the degree of entanglement in transmission from disordered media have been the subject of several investigations [9]-[14]. Since entangled photons are often exploited for encoding qubits, understanding their scattering properties in systems such as communication channels and biological samples that may exhibit disorder is of high importance for secure quantum communication and quantum imaging applications. Deleterious scattering can be a limiting factor and recently efforts have been made to compensate for these processes by classical control and feedback methods [15].

Experimentally, spatial quantum correlation has been measured for multiply scattered squeezed light [16] and the transport of quantum noise has been investigated [17]. Also, angular photon correlation has been captured in the transmission of super-Poissonian photon statistics through a multiple-scattering medium, showing a strong correlation between well-separated angular modes [18]. Two-photon speckle patterns, the analogue of classical speckle patterns for biphoton states, have been experimentally observed [19], [20] and their statistical distribution has been investigated theoretically [12]. In addition, recently, enhanced correlations have been experimentally observed within an angular range in the scattering of entangled photons from a dynamic complex medium with analogous interpretations to the interference effects in CBS. [14]. Furthermore, quantum correlations have been studied in the scattering of Gaussian states, such as thermal and squeezed-vacuum states, from disordered media using random-matrix theory and Monte Carlo simulation [13]. It has been shown that, in contrast to non-Gaussian states such as products or superpositions of number states, Gaussian states in the illumination always result in non-negative quantum correlations between two distinct scattered modes [13].

However, studies of the interaction of quantized fields with disordered geometries have mainly focused on the scattering of products of number states [9], [10] or the superposition of a few number state products [13]. Thus, one may certainly wonder how the scattering properties from mixed states or pure states consisting of superpositions of many photon pairs would differ. Additionally, quantum correlations have almost always been considered for different scattering modes in transmission from disordered media. It is of great interest to examine these correlations for coinciding angular directions, particularly recognizing that the probability for the detection of two photons in the same outgoing mode is quantifiable through two-photon absorption experiments [21].

In this paper, we calculate the angular quantum correlations of multiply scattered light, with a particular focus on two-photon states with varying degrees of entanglement and on the special case of backscattering. Our approach is to calculate the wave propagation for different realizations of point scatterers, and subsequently calculate the quantum correlations for different input quantum states of the light. Specifically, we solve the general Lippmann-Schwinger equation for scalar fields in conjunction with postprocessing to calculate quantum correlations. The results are then averaged over many realizations of the disorder. In our analysis, we made no assumptions (such as independent Gaussian distributions) on the statistical behavior of the scattering matrix elements. This provides several advantages over other commonly used approaches such as random-matrix theory. For example, the random-matrix theory can only capture the two-fold enhancement in the intensity of the CBS cone through the circular orthogonal ensemble assumption, yet it fails to reproduce the angular shape and the width of the CBS cone [22].

We focus our investigation on two categories of incident two-photon states: pure states in the form of coherent superpositions of jointly measurable photon pairs, possessing the maximum degree of entanglement, and their counterpart of fully mixed states with incoherent superpositions among the same group of paired single-photon states. The explicit distinctions between these excitations in terms of entanglement and the nature of superposition allow us to study their effects on the quantum correlation in backscattering. Additionally, we investigate the appearance of coherent two-photon backscattering in multiple scattering of these two distinct two-photon states. The effect of the degree of entanglement as well as the angular separation of jointly measurable twin photons on the quantum correlation and coherent two-photon backscattering are also studied. A key element in this subject is the interplay between induced and preserved correlations in multiple scattering.

The CBS effect in classical optics has been experimentally observed in multiple nonbiological disordered platforms including aqueous suspensions [1], [2], [23], powders [5], [24], cold atoms [25], multimode optical fibers [26], and amplifying random media [23]. This phenomenon is a consequence of the wave nature of the light and therefore has been observed for other forms of excitation such as acoustic waves [27], [28] and transient elastic waves [29] too. More recently, CBS cone measurement techniques have been also employed for the purpose of biological tissue characterizations. Since the full width of the enhanced backscatter cone is proportional to $\lambda/l^*$, in which $l^*$ is the mean free path length, CBS cone measurement is a powerful tool to characterize biological tissues and other materials, comparable with methods such as non-invasive diffuse reflectance measurement [30].

Multiple theoretical efforts towards modeling the line shape of CBS cone have been conducted by using diffusion theory [3], [4], [31]. In these studies, the contribution of maximally crossed diagrams [32] is added to the diffuse background in the ladder approximation in order to calculate the total averaged backscattered intensity within the diffusion approximation. Such maximally crossed diagrams represent the interference effects and result in the emergence of CBS peak, similar to Hikami vertices [33] which model weak localization corrections to the conductance in condensed-matter physics [34], and are responsible for a decrease in the diffusion coefficient in the multiple-scattering process. This weak localization effect, happening in weakly disordered geometries with $kl^* > 1$, is a precursor to strong localization effects such as Anderson localization [6], happening in strongly disordered geometries with $kl^* \leq 1$, in which $k = 2\pi/\lambda$ is the light wave number. The onset of the photon localization effects can be construed following the Ioffe-Regel criterion for the development of localization at $kl^* \simeq 1$, originally identified in localization of electron states [35].

The emergence of a weak localization effect is not always guaranteed in the presence of multiple scattering paths. Constructive interferences stem from the symmetry between reciprocal sequences of scattering events for classical waves [36]. Therefore, it is predictable that breaking the symmetry between time-reversed paths and introducing nonreciprocal phases interferes with this weak localization effect and suppresses the enhanced backscattered peak. It has been experimentally shown that by breaking the optical reciprocity using external magnetic fields and magneto-optical effects, the CBS effect can be suppressed [37], or be manipulated fully from weak localization to weak antilocalization [38]. Suppression of CBS in time-variant nonreciprocal systems, realized by ultrafast time modulation of a disordered medium, has also been observed [39]. The presence of nonlinearity similarly breaks down the reciprocity and transforms the CBS peak to a dip and again to weak antilocalization [40]. Therefore, the presence of multiple scattering events alone is not sufficient to observe CBS enhancement effects. Thus, a natural question that one may ask is how the presence of quantum fields, and in particular two-photon states in excitation, instead of coherent classical states —states produced by traditional lasers— may alter the emergence of CBS enhancement. Using a combination of theoretical and numerical tools, different aspects of this question are addressed in this work.

In the following, we start by presenting the theoretical model for calculating two-photon scattering in finite disordered configurations. We then study the quantum correlations in backscattering from multiple arrangements consisting of a few scattering centers. In the next step, the emergence of weak localization and coherent two-photon backscattering in the scattered two-photon currents is explored, along with the properties of the ensemble-averaged angular quantum correlation. We discuss the effect of disorder density and the order of ensemble averaging and demonstrate how the quantum correlation can be exploited as a probe for the dimensionality of entanglement for maximally entangled two-photon excitations. Analytical estimations for Gaussian statistics, quantum anticorrelation, and photon bunching effects are also discussed.

## II. Theoretical Model

The schematic layout of the optical configuration is displayed in Fig. 1. A two-photon input state (designated by the "8" sign) illuminates an ensemble of elastic point-like scatterers, shown as gray circles, which are randomly positioned within a rectangular cuboid of size $L_x \times L_y \times L_z$. The randomly positioned particles scatter the light into various output channels, generating single- and two- photon speckle patterns. In the backscattering direction these patterns are captured by two detectors and a coincidence counting circuit in the far field of the scattering medium.

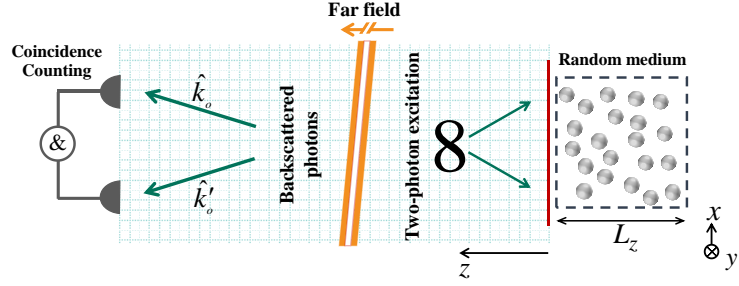

FIG. 1. Conceptual schematics of the setup. The input two-photon state, indicated by the "8" sign, illuminates the disordered geometry, and the backscattered photons are collected by two detectors at spatial modes corresponding to unit vectors $\hat{k}_o$ and $\hat{k}'_o$ in the far field region and studied in a coincidence counting circuit.

We study the effect of quantum entanglement and correlations of the biphoton excitation on the scattered single- and two- photon currents in the far field. We also investigate angular correlations in the far-field region for various deterministic and randomly distributed configurations, including properties of enhanced two-photon backscattering cones. Specifically, we consider two distinct monochromatic two-photon states for excitation. The first input state is a maximally entangled bipartite pure state with Schmidt rank $M$

$$|\Psi_M\rangle = M^{-1/2} \sum_{m=1}^{M} \hat{a}^\dagger_{\mathbf{q}_m} \hat{a}^\dagger_{-\mathbf{q}_m} |0\rangle \quad , \quad \hat{\rho}_{pure} = |\Psi_M\rangle\langle\Psi_M|. \tag{1}$$

The state is a superposition of photon pairs with direct anticorrelation between the transverse component of the wave vector of the first photon (i.e., $\mathbf{q}_m$) and the second photon (i.e., $-\mathbf{q}_m$). Multiple experimental implementations such as Spontaneous Parametric down-conversion (SPDC) processes in nonlinear crystals can realize such multi-dimensional entanglement between wavevectors of the generated signal and idler photons [41]-[43]. The Schmidt number of these entangled states may be determined with full state tomography or simply via near and far field intensity measurements exploiting the connection between coherent-mode decomposition in classical theory and Schmidt decomposition in quantum theory [44].

The second input state is the fully mixed counterpart of the entangled state, i.e., it is in the form of the averaged outer product of the corresponding state vectors with the same form of anticorrelation between transverse wave vectors of the photons

$$\hat{\rho}_{\text{mixed}} = M^{-1} \sum_{m=1}^{M} \hat{a}^{\dagger}_{\mathbf{q}_m} \hat{a}^{\dagger}_{-\mathbf{q}_m} |0\rangle\langle 0| \hat{a}_{\mathbf{q}_m} \hat{a}_{-\mathbf{q}_m}. \tag{2}$$

The photon creation and annihilation operators in quantization of electromagnetic field in transverse wave vector are denoted by $\hat{a}^{\dagger}_{\mathbf{q}_m}$ and $\hat{a}_{\mathbf{q}_m}$, respectively; with the state purity of $\text{Tr}[\hat{\rho}^2_{\text{mixed}}] = 1/M$, as expected. Except for the specified occupied modes in Eqs. (1) and (2), all the other incoming modes are in the vacuum states. In the following, these two-photon density operators are employed as excitation (see Fig. 1) in backscattering from a finite multiple-scattering medium.

Assuming perfect photodetection efficiency, the single- and two- photon currents, which may also be referred to as first and second order angular intensity correlation functions, can be obtained as

$$I_1(\mathbf{k}_o) = \text{Tr}[\hat{\rho}_{\text{out}} \hat{a}^{\dagger}_{\mathbf{k}_o} \hat{a}_{\mathbf{k}_o}], \tag{3}$$

and

$$I_2(\mathbf{k}_o, \mathbf{k}'_o) = \text{Tr}[\hat{\rho}_{\text{out}} \hat{a}^{\dagger}_{\mathbf{k}_o} \hat{a}^{\dagger}_{\mathbf{k}'_o} \hat{a}_{\mathbf{k}'_o} \hat{a}_{\mathbf{k}_o}], \tag{4}$$

which are proportional to the far field single-photon count rate at an observation direction $\mathbf{k}_o$ and the coincidence count rate at observation directions $\mathbf{k}_o$ and $\mathbf{k}'_o$ in two-photon detection, respectively [45]. Note that the detection joint probability in coincidence counting is related to the two-photon current as $P_2(\mathbf{k}_o, \mathbf{k}'_o) = (1 + \delta_{\mathbf{k}_o, \mathbf{k}'_o})^{-1} I_2(\mathbf{k}_o, \mathbf{k}'_o)$ [46]. The dimensionless ensemble-averaged quantum correlation function between two outgoing modes, ($\mathbf{k}_o$ and $\mathbf{k}'_o$), can then be expressed as

$$\overline{C(\mathbf{k}_o, \mathbf{k}'_o)} = \frac{\overline{I_2(\mathbf{k}_o, \mathbf{k}'_o)}}{\overline{I_1(\mathbf{k}_o)} \, \overline{I_1(\mathbf{k}'_o)}} = \frac{\overline{\text{Tr}[\hat{\rho}_{\text{out}} \hat{a}^{\dagger}_{\mathbf{k}_o} \hat{a}^{\dagger}_{\mathbf{k}'_o} \hat{a}_{\mathbf{k}'_o} \hat{a}_{\mathbf{k}_o}]}}{\overline{\text{Tr}[\hat{\rho}_{\text{out}} \hat{a}^{\dagger}_{\mathbf{k}_o} \hat{a}_{\mathbf{k}_o}]} \, \overline{\text{Tr}[\hat{\rho}_{\text{out}} \hat{a}^{\dagger}_{\mathbf{k}'_o} \hat{a}_{\mathbf{k}'_o}]}}. \tag{5}$$

Here, the overbars denote ensemble averaging over all realizations of the disordered medium (See Fig. 1). In a nonfluctuating disordered configuration (i.e., single realization) the scattered single- and biphoton currents are both fluctuating patterns referred to as speckle. The statistical quantum correlations buried in these fluctuations can be revealed by performing ensemble averaging. In the case of pure states, the numerator in Eq. (5) may equivalently be represented by $I_2(\mathbf{k}_o, \mathbf{k}'_o) = \langle :\hat{n}_{\mathbf{k}_o} \hat{n}_{\mathbf{k}'_o}: \rangle$ in which the brackets represent quantum expectation values in Dirac notation and $\hat{n}_{\mathbf{k}_o} = \hat{a}^\dagger_{\mathbf{k}_o} \hat{a}_{\mathbf{k}_o}$ is the occupation or particle number operator for output angular mode $\mathbf{k}_o$. The definition in Eq. (5) is similar to averaged quantum correlation functions explored in the literature; however, in several reported studies, the same terminology has been used for some slightly different forms of the averaged correlation functions [11], [16], [17]. In realistic experimental cases, the nonideal photodetector quantum efficiencies (QE) may be incorporated via a coefficient $K = \eta_2/\eta_1\eta_1$ with $\eta_1$ and $\eta_2$ representing the single- and two-photon QEs. In that case, the nonideal measured quantum correlation function is $\overline{C_{non-ideal}(\mathbf{k}_o, \mathbf{k}'_o)} = K\overline{C(\mathbf{k}_o, \mathbf{k}'_o)}$.

The multiple-scattering process can be modeled through a relation between incident and backscattered photon annihilation operators. The annihilation operator of the quantized electromagnetic fields for the backscattered field at output angular mode $\mathbf{k}_o$, i.e., $\hat{a}_{\mathbf{k}_o}$, is related to the annihilation operators of all the input fields via complex scattering matrix elements, as written in Eq. (6). Note that no Langevin noise operators are required for the lossless scattering process considered here, and the input and output (i.e., backscattered) creation and annihilation operators satisfy bosonic commutation relations,

$$\hat{a}_{\mathbf{k}_o} = \sum_{m=1}^{M} S_{\mathbf{k}_o, \mathbf{q}_m} \hat{a}_{\mathbf{q}_m} + \sum_{m=1}^{M} S_{\mathbf{k}_o, -\mathbf{q}_m} \hat{a}_{-\mathbf{q}_m}. \tag{6}$$

Equations (4) and (6) relate the complex scattering amplitudes derived from analysis of classical propagation (Maxwell's equations) to the coincidence counts in two outgoing modes. For random media or open cavities with random structure, it is difficult to determine the modes inside the medium. Still, such relation can be retrieved, connecting the input and output modes in the absence of absorption and amplification [8], [47]. Using mode-expansion (Eq. (6)), and simulation of classical scattering can significantly simplify the quantum-optical studies. Note that the second order intensity correlation function with $\mathbf{k}_o = \mathbf{k}'_o$ is proportional to the probability for the detection of both incident photons in the same outgoing mode, which is directly measurable via two-photon absorption experiments. Similarly, Eqs. (3) and (6) characterize an analogous connection between the scattering parameters and the single-photon currents or average number of photons in a single angular mode.

Using Eq. (6), relating the input and output bosonic operators in the scattering process, and the bosonic commutation relations (i.e., $[\hat{a}_{\mathbf{k}},\hat{a}_{\mathbf{k}'}]=[\hat{a}_{\mathbf{k}}^{\dagger},\hat{a}_{\mathbf{k}'}^{\dagger}]=0$ and $[\hat{a}_{\mathbf{k}},\hat{a}_{\mathbf{k}'}^{\dagger}]=\delta_{\mathbf{k},\mathbf{k}'}$, $\forall \mathbf{k},\mathbf{k}'$), the angular quantum correlation function for the maximally entangled state (Eq. (1)) and the fully mixed state (Eq. (2)) in coinciding observation directions are equal to $\overline{C_e(\mathbf{k}_o,\mathbf{k}_o)}$ and $\overline{C_m(\mathbf{k}_o,\mathbf{k}_o)}$, respectively

$$\overline{C_e(\mathbf{k}_o,\mathbf{k}_o)} = \frac{4M\overline{\left|\sum_{m=1}^{M} S_{\mathbf{k}_o,\mathbf{q}_m} S_{\mathbf{k}_o,-\mathbf{q}_m}\right|^2}}{\left(\sum_{m=1}^{M}\left(\left|S_{\mathbf{k}_o,\mathbf{q}_m}\right|^2 + \left|S_{\mathbf{k}_o,-\mathbf{q}_m}\right|^2\right)\right)^2}, \quad (7)$$

and

$$\overline{C_m(\mathbf{k}_o,\mathbf{k}_o)} = \frac{4M\overline{\sum_{m=1}^{M}\left(\left|S_{\mathbf{k}_o,\mathbf{q}_m}\right|^2 \left|S_{\mathbf{k}_o,-\mathbf{q}_m}\right|^2\right)}}{\left(\sum_{m=1}^{M}\left(\left|S_{\mathbf{k}_o,\mathbf{q}_m}\right|^2 + \left|S_{\mathbf{k}_o,-\mathbf{q}_m}\right|^2\right)\right)^2}. \quad (8)$$

For any arbitrary multimode coherent state, $|\{\alpha\}\rangle = \prod_{\mathbf{k}_c}|\alpha_{\mathbf{k}_c}\rangle$, the quantum correlation function, Eq. (5), is equal to one. This is due to the fact that the normally ordered expectation $I_2(\mathbf{k}_o,\mathbf{k}'_o)$ in this case is a product of quantum expectation values $\langle \hat{n}_{\mathbf{k}_o}\rangle$ and $\langle \hat{n}_{\mathbf{k}'_o}\rangle$, with parameters $\alpha_{\mathbf{k}_c}$ representing the amplitudes of modes in an ordinary mode expansion of the fields in the solution of classical Maxwell's equations. In other words, the two scattered modes are always uncorrelated for any input multimode coherent state.

In this work, the calculation of currents and correlation functions are implemented numerically for specific realizations of collections of point scatterers. We consider two classes of problems: (i) a small number of scatterers with specified spatial configurations, and (ii) a large number of particles which are uniformly and randomly distributed in all three dimensions. Specifically, in this case, the multiple scattering effect is studied for a collection of $N_p$ small particles which are randomly positioned within an imaginary cube with edge size $L = 20\lambda_0$ centered at the origin. Recently, we reported a scalar electromagnetic wave analysis in finite, wavelength-sized, randomized configurations to study the emergence of the CBS effect in multiply scattered light [48]. Here, we adapt a similar classical implementation, which calculates the scattering matrix elements in the far field domain (see Eq. (6)). Since the particles are much smaller than the wavelength, the analysis is isotropic and is in the Rayleigh scattering regime.

Specifically, we solve the general Lippmann-Schwinger equation for scalar fields in scattering from compositions of $N_p$ sub-wavelength scatterers with optical refractive index $n=1.5$ and radius $r=\lambda_0/2\pi$ and the scattering amplitudes are extracted in the far field region. In the case of randomized compositions, the multiple scattering depends upon the density of scatterers, defined as the ratio of the total volume occupied by scatterers to the volume of the cube, $\rho = 4\pi r^3 N_p/3L^3$. In such cases the effect of density on the observed coherent two-photon backscattering will also be examined. Once the scattering amplitudes between input and output modes are determined, Eqs. (3), (4), (7), and (8) are used to derive the single- and two-photon currents and normalized correlation function.

## III. Quantum Correlations in Backscattering from Few Scatterers

Prior to studying the interaction of two-photon states with random systems, it is useful to consider scattering from a small number (2-4) of particles with specific spatial locations, and particularly the significance of two-photon interference. Figure 2 illustrates the angular photon quantum correlation function $C(\mathbf{k}_o, \mathbf{k}'_o)$ in backscattering for few numbers of particles at $\mathbf{k}_o = \mathbf{k}'_o$ plotted vs the backscattering angle. The arrangement of scatterers in each analysis is shown in the inset of each panel. The spatial separation between the particles is set to $d=0.8\lambda$ and the vertical direction corresponds to the x axis (e.g., in Fig. 2(a) the two particles are positioned symmetrically on the x axis). Numerical results obtained upon illumination with maximally entangled and fully mixed states are shown with blue and red curves, respectively. The angle of incidence for the fields are symmetrically chosen around $\theta_{middle}=180°$ with $\Delta\theta = 5°$ as the separation angle and the Schmidt rank $M=2$. Note that $\theta_{middle}=180°$ incident angle corresponds to normal incidence with the wave propagating along -z direction (See Fig. 1). The plots correspond to backscattered angles, covering the entire half plane, $-\pi/2 \leq \theta \leq \pi/2$ (with *θ=0* corresponding to direct backscattering) and $\varphi=0$.

For any value of $M$ larger than one, the two-photon states in Eqs. (1) and (2) are dissimilar with respect to superposition and entanglement. The mixed state is an incoherent superposition of a group of paired single-photon states along anticorrelated directions, while the maximally entangled pure state is a coherent superposition of photon pairs. Upon scattering, each photon pair in the superpositions has multiple indistinguishable paths through the medium contributing to the two-photon detection at a given external angle. This results in an interference between the two-photon amplitudes, or in other words an interference of each photon pair with itself, in the joint detection scheme. These propagation paths can be understood by studying single-photon scattering from these geometries. The two-photon interferences along these paths add incoherently for input mixed states (i.e., Eq. (2) for $M>1$). In contrast, for maximally entangled pure states, the coherent superposition results in interference between photon pairs as well, causing a different collective behavior as measured by $C(\mathbf{k}_o,\mathbf{k}_o)$. It is noteworthy that while the nature of correlation in the mixed state is classical, the two-photon interference for these states results in quantum correlation behaviors dissimilar to input coherent states.

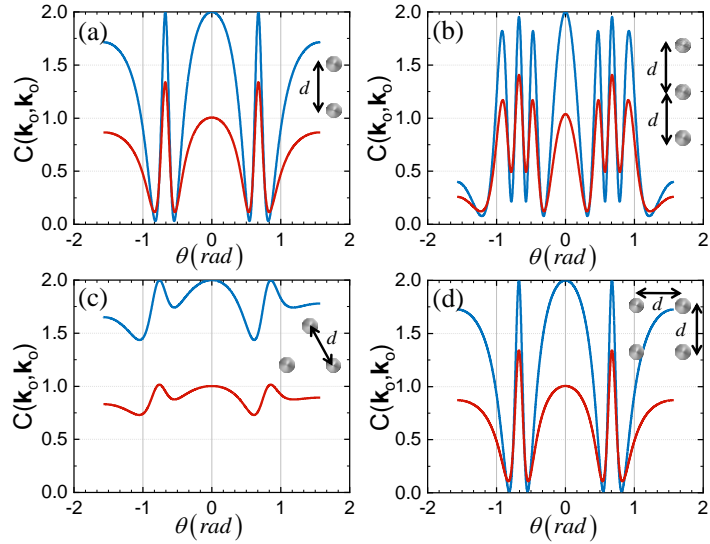

FIG. 2. The angular photon quantum correlation function $C(\mathbf{k}_o,\mathbf{k}'_o)$ in backscattering for few numbers of particles at $\mathbf{k}_o = \mathbf{k}'_o$ plotted vs the backscattering angle. The inset of figures indicates the associated arrangement of scatterers. Blue curves show correlation functions for maximally entangled pure input states and red curves show the correlation function for the fully mixed states. $M=2$, $\theta_{middle}=180°$, and $\Delta\theta=5°$ in both two-photon input states. The spatial separation between particles in all parts is $d=0.8\lambda$.

For both types of input states, the interference between two-photon amplitudes gives rise to oscillatory characteristics of the quantum correlation vs angle. This behavior can be clearly seen in Fig. 2. As one might expect, the shape of the correlation is very similar for the two cases, since they both correspond to interferences for two-photon amplitudes, but *the contrast of these oscillations is larger for entangled pure states*. The quantum correlation reaches a maximum value of $M$ for pure states (see also Fig. 3) and to a smaller value for mixed states that depends on the arrangement of scatterers and the separation angle for the mixed states.

To interpret these results, it is useful to first note that the two-photon currents and correlation functions of the excitation are essentially delta functions $\delta(\mathbf{q}_m, -\mathbf{q}_m)$: for the incident light, if a photon detector at angle $\mathbf{q}_m$ registers a photon, then the second detector will register a photon precisely at angle $-\mathbf{q}_m$, and the photon pairs will never produce joint photodetection events for detectors located at the same angle. On the other hand, after scattering from the medium, as can be seen in Fig. 2 (see also Fig. 3), the correlation functions for coinciding observation angles can have non-zero values for both pure and mixed states and also reach their maximum values at certain backscattered angles for pure states.

For the case of two scatterers, the angular positions of these maxima and minima can be predicted considering coherent superpositions and two-photon interference in the framework of a double slit problem. To further examine the effect of Schmidt rank and separation angles on the correlations, we repeat the analysis for the two-scatterer configuration with spatial separation $d = \lambda$, Schmidt ranks $M = 2, 5, 8$, and separation angles $\Delta\theta = 2°, 10°$. The illumination angles are again located around $\theta_{middle} = 180°$ (normal to the line connecting the two particles and along the -z direction) and the results are plotted in Fig. 3.

An immediate conclusion from Figs. 2 and 3 is that *the number of maxima and minima in the correlation pattern depends only on the number and arrangement of the scatterers and does not depend on the number of (coherently or incoherently) superposed photon pairs.* Changing the number of photon pairs in the superposition modifies the maximum correlation and slightly shifts the angular location of the dips in the patterns. Also, we see higher oscillation amplitude in the correlation patterns from maximally entangled states compared to fully mixed state, which is consistent with the results and discussion of Fig. 2. Additionally, by increasing the number of pairs in the superposition, the peaks (corresponding to constructive interference) broaden. In particular, for small separation angles, e.g., $\Delta\theta = 2°$ as shown in parts (a) and (c), very sharp peaks can be observed, while having contributions from more pairs widens the peak. These sharp peaks are direct consequences of two-photon interference within each photon pair in the two-photon states, and as the result can be seen for both mixed states with statistical mixture and entangled pure states with coherent superpositions. For small separation angles, the twin photons within each pair excite the two scatterers at very close angles and thus result in quite similar possible paths for two-photon amplitudes; therefore, the interference between these two-photon amplitudes lead to fast angular oscillations with sharp peaks.

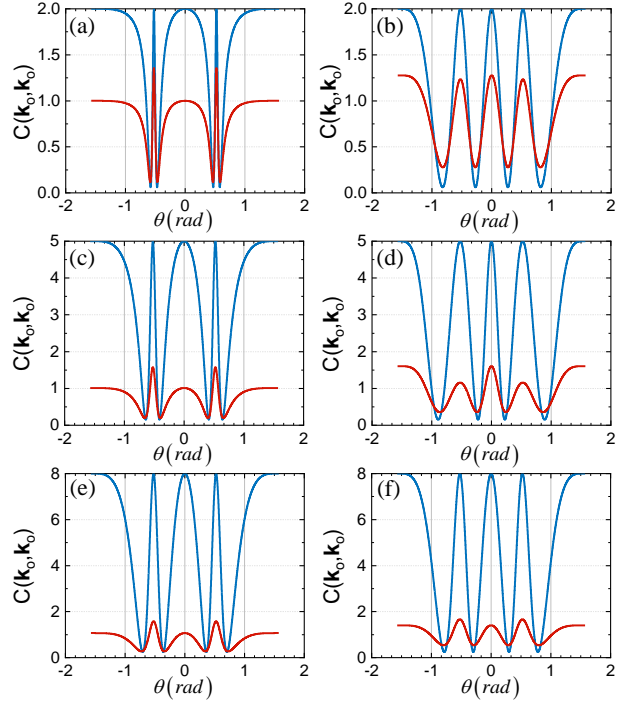

FIG. 3. The angular photon quantum correlation function $C(\mathbf{k}_o, \mathbf{k}_o)$ in backscattering from two particles plotted vs the backscattering angle. Particles are symmetrically positioned on x axis with spatial separation $d = \lambda$. Blue curves show correlation function for maximally entangled pure input states and red curves show the correlation function for the fully mixed states. For both two-photon input states, $\theta_{middle} = 180°$ and the Schmidt rank is $M = 2$ (a,b), $M = 5$ (c,d), and $M = 8$ (e,f). The separation angle in incident states is $\Delta\theta = 2°$ (a,c,e) and $\Delta\theta = 10°$ (b,d,f).

## IV. Quantum correlations in backscattering from random media

### IV.1. Coherent backscattering in the two-photon current

We now turn to the consideration of scattering from random distributions of point scatterers that are sufficiently dense to exhibit such multiple scattering. Maximally entangled, fully mixed, and classical coherent input states are considered in this study. Similar to reported studies on few scatterers, the necessary far field scalar responses are calculated for a composition of uniformly distributed particles and for input waves with distinct tilt angles, corresponding to components in Eqs. (1) and (2). To carefully separate the coherent effect of specular reflection, which also survives the ensemble averaging [48], $\theta_{middle}$ is set to $140°$ (corresponding to an angle of 40 degrees relative to the normal of the input facet of the cube), so that the emergence of CBS and specular reflection can be inspected independently at different directions.

In Fig. 4 the calculated normalized two-photon current is depicted in scattering from random clusters of $N_p = 5000$ particles in a box with edge size $L = 20\lambda_0$, ensemble-averaged over $N_r = 1000$ distinct realizations of the disorder. The black and yellow curves show the angular dependance of the two-photon current for maximally entangled pure and fully mixed states, respectively. The incident waves (in Eq. (1) and (2)) are symmetrically chosen around $\theta_{middle} = 140°$ with the separation angle $\Delta\theta = 1°$ and Schmidt ranks $M = 1, 2, 5, 10$ for parts (a)-(d). To get a clearer comparison with traditional CBS curves, all curves here are scaled to have a maximum of two.

The Schmidt rank $M$, corresponds to $M$ jointly measurable photon pairs in the illumination and hence $2M$ distinct angles of incidence involved in the two-photon excitation process. The blue curves show the customary classical CBS observed for a single plane wave excitation of the same geometry with the input field direction aligned with one of these illumination angles. The chosen angle here is $\theta_{middle} + \Delta\theta$ and the enhancement factor is found to be 1.63. The red curves show the two-photon current for the same classical single wave excitation. To further examine the interference effects between the components in the superposition, the cyan curves show the currents for the same geometries and incoherent addition of the contributions from all $2M$ incident angles with plane wave coherent state excitation in each angle. The plots cover the entire backscattered half plane, $-\pi/2 \leq \theta \leq \pi/2$. The backscattered and specular reflections may be respectively observed at $\theta_{CBS} = -40° \approx -0.7 [rad]$ and $\theta_{Specular} = 40° \approx 0.7 [rad]$.

The coherent two-photon backscattering effect can be observed in all cases confirming the presence of weak localization and constructive interference between two-photon amplitudes in reciprocal paths. Another observation is that the degree of enhancement in all shown cases is less than two (for customary CBS in blue curves) and less than four (for all other curves). The observation of an enhancement less than two in conventional CBS is a known consequence of the finite size of the disordered medium in all three dimensions, and has been previously predicted and studied [48]-[50].

The width of the observed CBS cone in the backscattered single-photon current for a classical source (blue curve) is $FWHM \approx 0.1539 [rad]$ and the mean free path length in multiple scattering can be extracted from this width to be around $l^* = 0.689\lambda$. Thus, multiple scattering is occurring in a weakly disordered structure with $kl^* \approx 4.33 > 1$ in harmony with the Ioffe-Regel criterion [35] for the appearance of localization at $kl^* \simeq 1$. It is important to note that the angular line shape and width of the enhanced backscattering cone strongly depends on the finiteness of the geometry in all three dimensions. Although in most initial theoretical investigations of multiple scattering in the mesoscopic regime the finite length of the sample and the finite size of the excitation beam were neglected, these finite parameters can play an appreciable part in actual experimental studies. The finite length of the medium eliminates the possibility of long scattering pathlengths. Finite-size effects within the diffusion approximation with two absorbing boundary conditions has been utilized to show theoretically that there exists a direct relationship between the length of photon trajectories and the width of the CBS cone [51]. In a scalar analysis it was shown that reducing the width of the disordered slab results in shorter possible photon loops and hence wider cones (as is the case here). The effect of finite thickness of the sample on producing wider cones with lower peak values has also been observed experimentally [50]. This interpretation also agrees with picosecond time-resolved measurements of the peak profile in the case of femtosecond laser pulse excitations of a semi-infinite random media [52]. Narrower peaks are observed at larger time delays corresponding to photons traveling longer distances in the medium. The finite transverse extent of the excitation beam also broadens the peak and reduces the ratio of enhancement by excluding the contribution of time-reversed paths with the distance between the first and last scattering event longer than the size of the beam [49]). Accordingly, the finiteness of the analyzed geometry in all three dimensions here produces broader CBS peaks with the enhancement ratio less than two.

The full width at half maximum for the two-photon currents are different from that of the one-photon current. It is $FWHM \approx 0.1304[rad]$ for the classical single-wave coherent input state and varies for the two-photon input states depending on the Schmidt rank and angle of separation, as can be seen in Fig. 4. By increasing the Schmidt rank, the number of photon pairs in the superposition increases accordingly and hence a broader range of incident angles (and therefore CBS angles) are excited. As the result, broader coherent two-photon backscattering cones are observed, e.g., in Fig.4 (d). The ratio of peak value to the background signal (incoherent albedo) is one of the important quantities to investigate in the weak localization effects. Here, the observed degree of enhancements varies from 2.40 to 1.81 for two-photon currents by increasing the Schmidt rank.

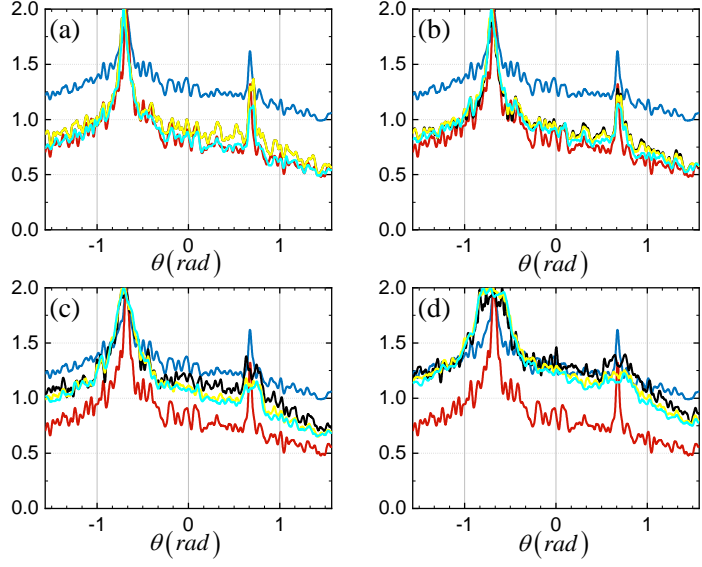

FIG. 4. The normalized ensemble-averaged angular two-photon current $\overline{I_2(\mathbf{k}_o, \mathbf{k}_o)}$ in backscattering from a composition of $N_p = 5000$ point scatterers plotted vs the backscattering angle for maximally entangled pure two-photon states (black), fully mixed states (yellow), and their incoherently added classical equivalents (cyan). Blue curves show the single-photon current for the classical single wave excitation (i.e., customary CBS) and the red curve shows the corresponding two-photon current in that classical single wave excitation [see text for more details]. Angles $\theta_{CBS} \approx -0.7\,[\text{rad}]$ and $\theta_{Specular} \approx 0.7\,[\text{rad}]$ correspond to backscattering and specular reflections for middle incident angle, respectively. The separation angle in incident states is $\Delta\theta = 1°$ and the Schmidt rank is (a) $M = 1$, (b) $M = 2$, (c) $M = 5$, and (d) $M = 10$.

There is an evident resemblance in Fig. 4 between the currents for both types of two-photon input states and their incoherently added classical equivalent. This indicates *that the nature of the superposition in the two-photon source (coherent superposition vs statistical mixture) has almost no impact on the joint photon probability after ensemble averaging, although some differences can be seen for intermediate values of M* (e.g., in Fig. 4(c)). One may conclude that sufficient, strong-enough multiple scattering effects can dissolve the signature of those superpositions in measurements of the two-photon currents (i.e., coincidence counts in an experiment) in the weak localization regime.

### IV.2. Ensemble-averaged angular quantum correlation

While the two-photon current exhibits a coherent backscattering effect that is nearly independent of the nature of the superposition of the incident two-photon state, the ensemble-averaged quantum correlation function does exhibit important differences. Figure. 5 illustrates the ensemble averaged angular quantum correlation for the same disordered system for maximally entangled (blue) and fully mixed (red) states as the illumination sources.

For each realization of the randomly located scatterers, the single- and two-photon currents and consequently the quantum correlation function exhibit granular features referred to as single- and two-photon speckle patterns (further discussion of these speckle patterns is given in Appendix I). By performing ensemble averaging over these fluctuations the underlying statistical correlations can be extracted, shown in Fig. 5. The averaging over realizations for quantum correlation function here is performed as denoted in Eq. (5); however, this step can be conducted via alternative ways. Further discussions can be found in Appendix II on the order of ensemble averaging and its impact on the results.

As we saw in the discussion of scattering from a few particles in the previous section, we see that *multiple-scattering events can induce quantum correlations in the output modes*. The correlation functions for both forms of two-photon excitation are comprised of Kronecker delta functions with non-zero values corresponding to the angles of the twin photons. Thus, the joint detection probability and the two-photon current is always zero for coinciding detectors in the absence of the disordered structure. However, one can see that upon multiple scattering, the quantum correlation for the observation of both photons in the same mode at the output increases to nonzero values.

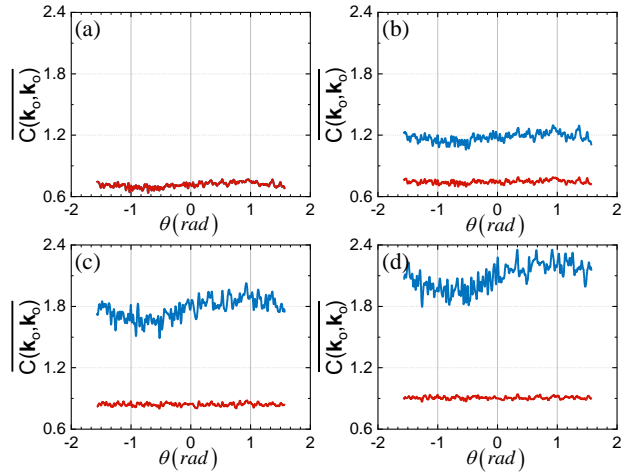

FIG. 5. The ensemble-averaged angular photon quantum correlation function $\overline{C(\mathbf{k}_o, \mathbf{k}_o)}$ in backscattering for the studied cases presented in Fig. 4. Blue curves show correlation functions for maximally entangled pure input states and red curves show the correlation functions for the fully mixed states.

These correlations are independent of the observation angle for the mixed state illumination but show a small variation with angle for the entangled pure state illumination, with a broad minimum around the backscatter direction and a small enhancement around the specular direction. Additionally, for the small separation angle studied here, *the induced correlations for the entangled pure states are substantially larger than the induced correlations by the fully mixed states, with the entanglement-induced correlation enhancement increasing with larger values of $M$*. The most striking property, though, is that *unlike the two-photon current, which exhibits a CBS peak, the quantum correlation function (which is a normalized quantity) does not*; in fact, it shows a slight *reduction* in magnitude around the backscatter direction relative to the diffuse background. To further examine the different behavior for pure and mixed input states, in the next analysis we explicitly focus on the impact of separation angle on the coherent two-photon backscattering line shape and the induced quantum correlations. The broad dip present in the correlation for the two-photon pure state around the CBS observation angle has no counterpart for mixed two-photon states and *is therefore a direct consequence of the interference between two-photon amplitudes of different photon pairs*.

### IV.3. Effect of angular separation: interference of two-photon amplitudes from different photon pairs

In the geometry of our problem, as illustrated in Fig. 1, photon pairs are incident on the random medium with a separation angle $\Delta\theta$. It is interesting to study the effect of this separation angle on the quantum correlation for illumination by a coherent superposition (Eq. (1)) or by an incoherent mixture (Eq. (2)). The results are shown in Figs. 6 and 7, where the ensemble-averaged angle-resolved two-photon currents and correlations are plotted for these input states. The specifications of the disordered system are identical to those reported for Figs. 4 and 5 and the input Schmidt ranks are fixed at $M = 2$, while studying the input separation angle $\Delta\theta$.

The behavior observed in Fig. 6 regarding the emergence of two-photon localization and its dependance on the input separation angle is similar to the results (Fig. 4) obtained for varying the Schmidt rank of the two-photon states. The same broadening effects and decrease in the ratio of enhancement can be seen by increasing the separation angle. Additionally, for large $\Delta\theta$ and with $M = 2$ the emergence of four distinct localization effects can be clearly seen in Fig.6 (d). Overall, one may conclude that the effects of Schmidt rank and $\Delta\theta$ act essentially equivalently.

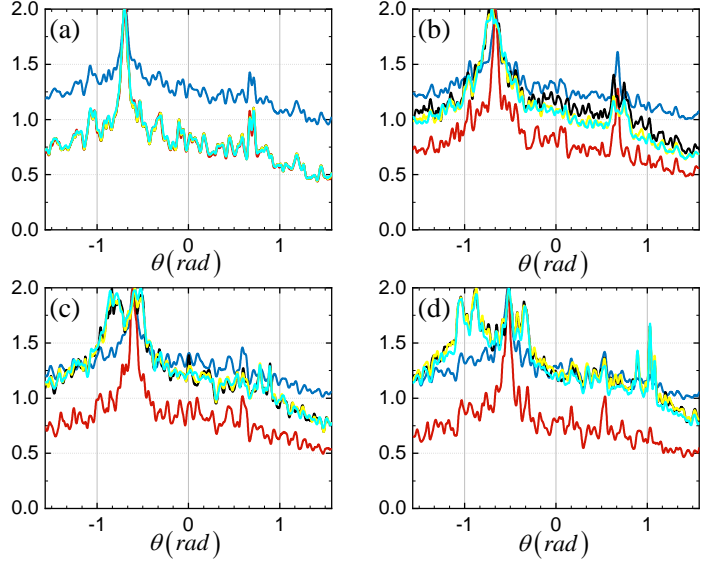

FIG. 6. The normalized ensemble-averaged angular two-photon current $\overline{I_2(\mathbf{k}_o, \mathbf{k}_o)}$ in backscattering from a composition of $N_p = 5000$ point scatterers plotted vs the backscattering angle for maximally entangled pure two-photon states (black), fully mixed states (yellow), and their incoherently added classical equivalents (cyan). Blue curves show the single-photon current for the classical single wave excitation (i.e., customary CBS) and the red curve shows the corresponding two-photon current in that classical single wave excitation. Angles $\theta_{CBS} \approx -0.7\,[\text{rad}]$ and $\theta_{Specular} \approx 0.7\,[\text{rad}]$ correspond to backscattering and specular reflections for middle incident angle, respectively. The Schmidt rank is $M = 2$ and the separation angle in incident states is (a) $\Delta\theta = 0.25°$, (b) $\Delta\theta = 2°$, (c) $\Delta\theta = 5°$, and (d) $\Delta\theta = 10°$.

An immediate conclusion from Fig. 7 is that *at large separation angles for the modes in the input two-photon states, the quantum correlations for entangled and mixed states are quite comparable and independent of the observation angle*. However, *distinct differences can be seen for the two cases when the separation angles are small and comparable to the speckle correlation angles*, which is around $\Delta\theta_{corr} = 2.25°$ for this composition (see Appendix III).

This can be understood considering the coherent superposition between photon pairs present in the pure state. *For small separation angles, it is possible for the two-photon amplitudes from different photon pairs to constructively interfere with each other as they experience similar scattering paths within the medium, and hence increase the quantum correlation in the scattered light compared to the mixed states.*

In the case of sufficient, statistically independent scattering events (occurring for large $\Delta\theta$), circular Gaussian statistics can be shown to accurately predict the statistical behavior of the geometry, and thus the quantum correlations.

See Appendix IV for details on utilizing complex Wick's theorem for statistical averaging of quantum correlation function, considering statistically independent scattering matrix elements

$$\overline{C_e\left(\mathbf{k}_o,\mathbf{k}_o\right)}\bigg|_{Gaussian} = \overline{C_m\left(\mathbf{k}_o,\mathbf{k}_o\right)}\bigg|_{Gaussian} = \frac{2M}{1+2M}. \qquad (9)$$

Consistently, the quantum correlation for the case studied here converges to the value 0.8 (see Figs. 7 (e) and (f)). As the ensemble-averaged correlation function depends on the degree of entanglement, this quantity, which is experimentally accessible through multiple techniques including two-photon absorption, can be used to determine the degree of entanglement, knowing that the illumination is pure. This is not the case for non-coinciding detectors (i.e., detectors at two different angles), as for those cases the averaged quantum correlation is always equal to one half, regardless of the value of $M$. Also, given the Gaussian assumptions, an anticorrelation can be clearly seen for all values of Schmidt rank for both two-photon input states. Clearly for small separation angles $\Delta\theta$ (Fig. 7) it is possible to see correlations greater than one for pure entangled states. See Appendix IV for further discussions regarding spatial photon bunching and quantum correlations between two distinct angular modes. It is noteworthy that the observed increase in correlation for small $\Delta\theta$ for entangled pure states is fundamentally different from the increased 2-channel correlation function in the Anderson localization regime [10].

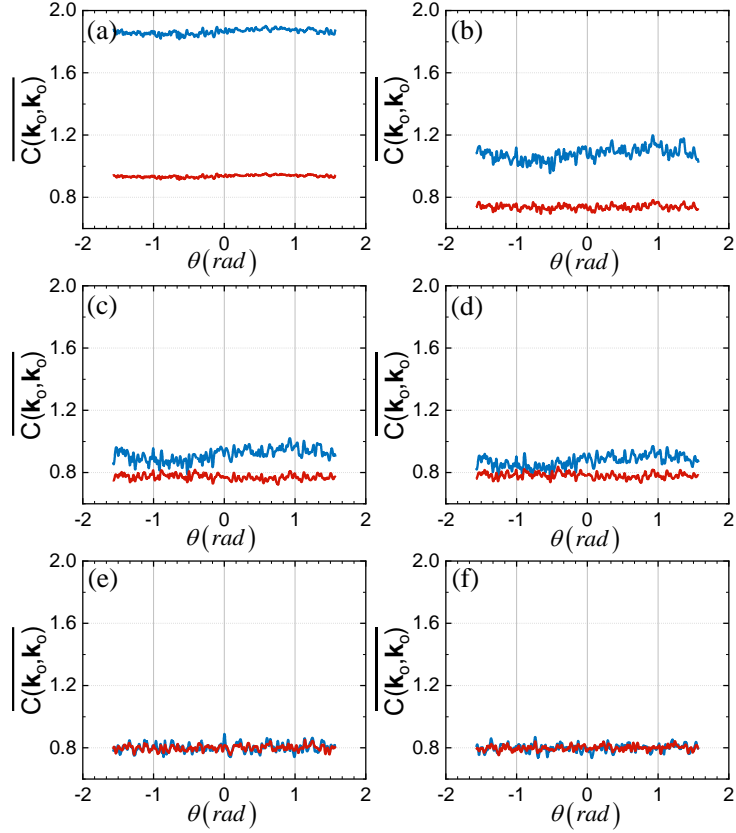

FIG. 7. The ensemble-averaged angular photon quantum correlation function $\overline{C(\mathbf{k}_o, \mathbf{k}_o)}$ in backscattering from a composition of $N_p = 5000$ point scatterers plotted vs the backscattering angle. Blue curves show correlation functions for maximally entangled pure input states and red curves show the correlation functions for the fully mixed states. The quantum correlation function is found to be independent of observation angles. The Schmidt rank is $M = 2$ and the separation angle in incident states is (a) $\Delta\theta = 0.25°$, (b) $\Delta\theta = 1.25°$, (c) $\Delta\theta = 1.8°$, (d) $\Delta\theta = 2°$, (e) $\Delta\theta = 5°$, and (f) $\Delta\theta = 10°$.

## V. Effect of Disorder Density and Numerical Convergence Study

The classical CBS line shape depends on the mean free path length within the disordered structure, or in other words the density of scattering centers [2]. Here, we examine the effect of density on the multiple scattering of two-photon states and in particular the coherent two-photon backscattering. Specifically, we study the normalized two-photon current in backscattering from $N_p = 3000, 6000, 10000$ particles in a box with edge size $L = 20\lambda_0$ (corresponding to the particle densities $\rho = 0.00633, 0.01267, 0.02111$) with ensemble-averaging over $N_r = 1000$ compositions of particles. The Schmidt rank is $M = 2$ and the angle of incidence for the fields are again symmetrically chosen around

$\theta_{middle} = 140°$ with separation angle $\Delta\theta = 1°$. Figure 8 provides a detailed look at the dependance of CBS and coherent two-photon backscattering cones on the density of disordered structures, verifying that the width of the enhancement cones increases by increasing the level of disorder (increasing density of particles), consistent with reduction of the mean free path length within the geometry, i.e., $width \propto \lambda/l^* \propto \rho$.

The degree of enhancement in all studied cases is less than two (for blue curve) and less than four (for all other curves) and is larger for denser media. Reduction of the mean free path length for a fixed box size effectively results in the possibility of longer reciprocal paths in the geometry, and hence taller enhancement peaks. Another relevant conclusion is that the line shapes of the coherent two-photon backscattering cones (black and yellow) and their incoherently added classical equivalent (cyan) converge to their classical counterpart (red) with single wave excitation in denser structures. The similarities between responses for two-photon states and their incoherently added classical equivalent was previously discussed (see discussions on Figs. 4 and 6). Additionally, here these responses are similar to the result for a classical single wave excitation for the case with more scattering particles. Note that with Schmidt rank $M = 2$, four waves with different incident angles contribute to the response, each producing a cone at their corresponding backscattering direction. Thus, for wider peaks in the event of denser disordered structures, the normalized incoherent summation of the currents (cyan curve) produces a peak wider but with enhancement quite close to the response from only one of the incident waves (red).

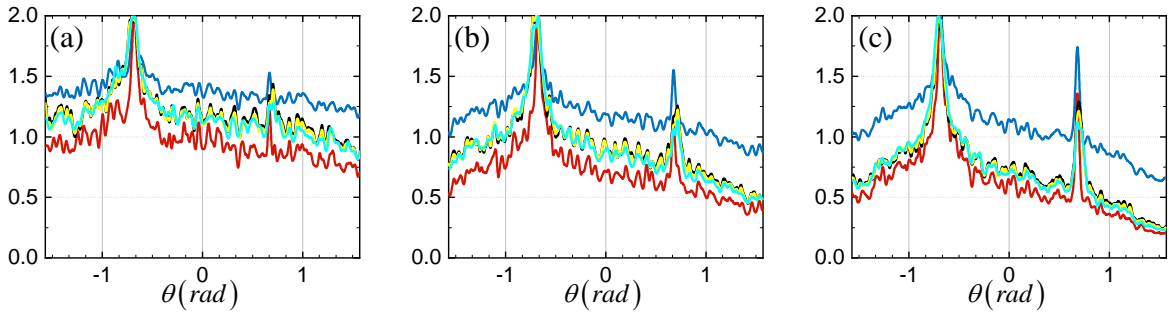

FIG. 8. The normalized ensemble-averaged angular two-photon current $\overline{I_2(\mathbf{k}_o, \mathbf{k}_o)}$ for different disorder densities plotted vs the backscattering angle for maximally entangled pure two-photon states (black), fully mixed states (yellow), and their incoherently added classical equivalents (cyan). Blue curves show the single-photon current for the classical single wave excitation (i.e., customary CBS) and the red curve shows the corresponding two-photon current in that classical single wave

excitation. The Schmidt rank is $M = 2$ and the separation angle in incident states is $\Delta\theta = 1°$ in scattering from (a) $N_p = 3000$, (b) $N_p = 6000$, and (c) $N_p = 10000$ particles.

Next, we examine the convergence behavior of ensemble averaging, i.e., the effect of the number of the realizations of the system, on the CBS line shape and angular correlations. These analyses are conducted by keeping all the other parameters unchanged, and the results are shown in Fig. 9. Part (a) shows the normalized two-photon current in backscattering from $N_p = 5000$ particles in a box with edge size $L = 20\lambda_0$ and with ensemble averaging over $N_r = 10000$ distinct clusters of particles. The angle of incidence for the fields are symmetrically chosen around $\theta_{middle} = 140°$ with Schmidt rank $M = 2$ and separation angle $\Delta\theta = 1°$. The color mapping for different incident fields is the same as the previously reported results (See Figs. 4,6, and 8). The comparison can be made with the Fig. 4(b) corresponding to the same set of parameters yet with ensemble averaging over much fewer number of realizations. One can clearly observe the cone width and height is essentially constant, with the same pattern of response for different states of excitation, and in overall a numerically converging behavior. Figure. 9(b) shows the angular quantum correlation for the same disordered system and the results for the two different number of realizations are plotted on top of each other. Blue and black (red and yellow) curves correspond to maximally entangled (fully mixed) two-photon states as illumination sources. The noisier curves (blue and red) are the responses for averaging over $N_r = 1000$ realizations and black and yellow curves show the same response while averaging over $N_r = 10000$ realizations. A similar numerically converging behavior may be observed for the correlation functions; and thus averaging over ensembles as large as $N_r = 1000$ is sufficient to capture the coherent effects within the multiple scattering schemes and notably between reciprocal paths.

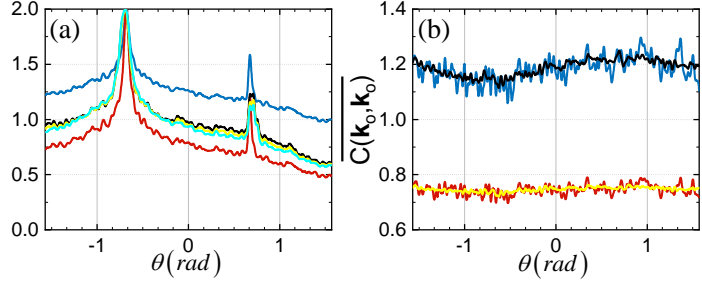

FIG. 9. Numerical convergence study. (a) The normalized ensemble-averaged two-photon current $\overline{I_2(\mathbf{k}_o,\mathbf{k}_o)}$ for the same studied case presented in Fig. 4(b) while averaging over $N_p = 10000$ distinct compositions of particles. (b) The ensemble-averaged photon quantum correlation function $\overline{C(\mathbf{k}_o,\mathbf{k}_o)}$ in backscattering for the studied case presented in Fig. 5(b); plotted in blue and red curves for $N_r = 1000$, and black and yellow curves for $N_r = 10000$.

## VI. Summary and Conclusion

In conclusion, utilizing a model of point-like elastic scatterers, we have investigated the two-photon current and angular two-photon quantum correlation function in backscattering from systems of a few particles and from a large number of randomly positioned particles in a finite volume. Studies are conducted for bipartite states in pure or mixed states and the role of entanglement, classical correlations, and quantum interference are examined. We demonstrate that, analogous to the classical CBS enhancement in the scattered intensity which is the well-known signature for weak localization of photons in multiple scattering, coherent two-photon backscattering emerges owing to constructive interference between reciprocal paths in multiple scattering of two-photon states of the light. The joint photon probability has almost no dependence on the nature of the superposition in the two-photon source (coherent superposition vs statistical mixture), and thus multiple scattering effects can dissolve the signature of those superpositions in measurements of the two-photon currents.

In contrast, the ensemble-averaged angular quantum correlation function (which is a normalized quantity) does exhibit significant differences for maximally entangled versus fully mixed illumination states. Interestingly, the averaged angular quantum correlation function does not show a narrow backscattered peak, but rather shows a slight *reduction* in magnitude in a broad region around the backscatter direction relative to the diffuse background for the two-photon pure state, but not for mixed two-photon states and thus is a direct consequence of the interference between two-photon amplitudes of different photon pairs. The possibility of using ensemble-averaged quantum correlations as a probe for the dimensionality of entanglement has been presented and the effect of particle density, photon anticorrelation, and bunching effects are also discussed. Studies of the dependance of the quantum correlations on the separation angle of the incident photon pairs show that, for small separation angles, it is possible for the two-photon amplitudes from different photon pairs in entangled states to constructively interfere with each other as they experience similar scattering paths within the medium, and hence increase the quantum correlation in the scattered light compared to the mixed states. At large separation angles, the correlations of entangled and mixed pair states are essentially similar.

**Appendix I: Granular Character of Quantum Correlation Functions:**

The single- and two-photon currents in backscattering from the geometries studied here exhibit granular angular character typically called speckle patterns. Therefore, the quantum correlation function corresponding to each cluster has a granular angular nature. The angular dependance of the correlation function from a sample single cluster is presented in Fig. 10. Parts (a) and (c) show the correlation for a cluster of $N_p = 5000$ scatterers illuminated with a maximally entangled pure state with Schmidt rank $M = 2$, and parts (b) and (d) show this correlation for the counterpart fully mixed state. Here, the plane waves' incidence angles are symmetrically chosen around $\theta_{middle} = 140°$ with $\Delta\theta = 0.75°$ (parts (a) and (b)) and $\Delta\theta = 10°$ (parts (c) and (d)) as the separation angle. Similar to the results obtained for few-particle scattering (see Figs. 2 and 3), the range of oscillations (related to the visibility of the speckle patterns) in the correlation is always larger for pure entangled states. This characteristic is closely connected to two-photon visibility in backscattering. This can be understood considering the distinct nature of the photon pair superpositions in the fully mixed and entangled pure states. The two-photon interferences occur between photon pairs and within pairs for the input pure state, while the two-photon interferences within pairs (between associated two-photon amplitudes of each pair) are added statistically to produce the result in the mixed state.

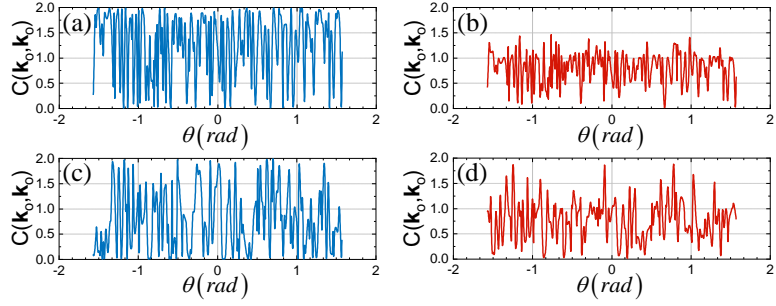

FIG. 10. The angular photon quantum correlation function $C(\mathbf{k}_o, \mathbf{k}_o)$ in backscattering from a composition of $N_p = 5000$ point scatterers plotted vs the backscattering angle for (a,c) maximally entangled pure input states and (b,d) fully mixed states. For both two-photon input states, $\theta_{middle} = 140°$, Schmidt rank is $M = 2$, and the separation angle in incident states is $\Delta\theta = 0.75°$ (a,b) and $\Delta\theta = 10°$ (c,d).

**Appendix II: The order of Ensemble Averaging**

In most of the previously reported studies on the quantum correlation function in disordered geometries as well as the results presented here, the ensemble averaging is conducted by finding the statistical average for the numerator and the denominator of the correlation function (see Eq. (5)) separately. As discussed in section II, in some other reports the average values for the single-photon currents contributing to the denominator are obtained independently in the process. In numerical investigations (such as this paper) as well as experimental studies it is possible to use a different form of ensemble-averaged correlation by averaging over correlation functions calculated separately for each realization as

$$\overline{C'(\mathbf{k}_o, \mathbf{k}'_o)} = \overline{C_{(i)}(\mathbf{k}_o, \mathbf{k}_o)} = \overline{\left(\frac{I_2(\mathbf{k}_o, \mathbf{k}'_o)}{I_1(\mathbf{k}_o) I_1(\mathbf{k}'_o)}\right)}, \quad (10)$$

in which $C_{(i)}(\mathbf{k}_o, \mathbf{k}_o)$ is the quantum correlation function calculated separately for the $i$ th realization. The $\overline{C(\mathbf{k}_o, \mathbf{k}'_o)}$ function estimates how the averaged two-photon current is related to the averaged multiplication of single-photon currents; with a view to study if the observation of photons, on average, in the joint detection scheme is correlated, anticorrelated, or uncorrelated. On the other hand, $\overline{C'(\mathbf{k}_o, \mathbf{k}'_o)}$ function, assesses similar type of correlation in joint detection for each realization and then finds the average correlation behavior. As one can clearly see, these two quantities are closely related. Figure. 11 depicts these two correlation functions at coinciding observation angles for

$N_r = 1000$ set of disordered realizations of $N_p = 5000$ point-like particles positioned in a box of size $L = 20\lambda_0$ for maximally entangled (blue) and fully mixed (red) states with $M = 2$, $\theta_{middle} = 140°$, and separation angles $\Delta\theta = 0.5°$ (a,c) and $\Delta\theta = 8°$ (b,d).

One may notice that for small and large separation angles, these two functions are both almost independent of the observation angle, yet with some differences between them. At small separation angles, the responses for the mixed states are quite similar; however, as we saw before, the constructive interference between the two-photon amplitudes associated with different photon pairs leads to larger correlations for pure entangled input states. This distinction is smaller for $\overline{C'(\mathbf{k}_o,\mathbf{k}_o)}$ function compared to $\overline{C(\mathbf{k}_o,\mathbf{k}_o)}$. On the other hand, at large separation angles, these two averaged functions converge to the same value with a slightly larger statistical variance for $\overline{C(\mathbf{k}_o,\mathbf{k}_o)}$ function. As discussed before, for large separation angles the correlation functions converge to the analytical values derived using complex Gaussian statistics, i.e., to the value $0.8$ for the case discussed here. See Appendix IV for the relevant calculations and discussions.

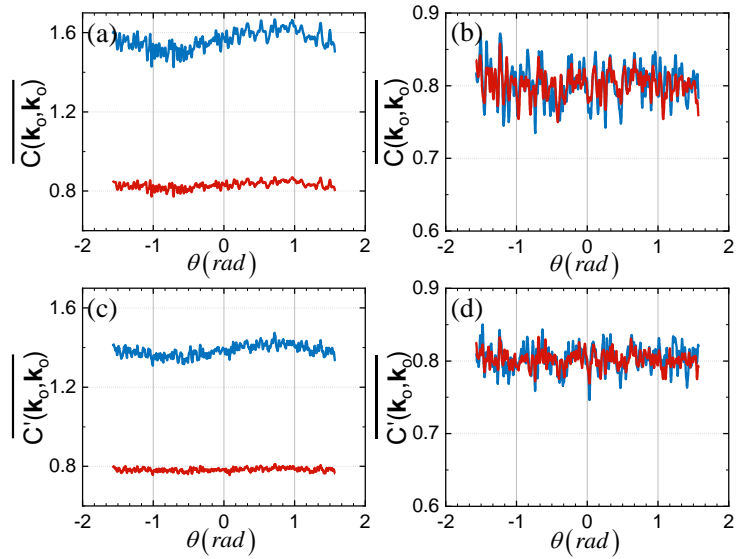

FIG. 11. Study on the order of ensemble averaging. The ensemble-averaged photon quantum correlation functions $\overline{C(\mathbf{k}_o,\mathbf{k}_o)}$ (a,b) and $\overline{C'(\mathbf{k}_o,\mathbf{k}_o)}$ (c,d) in backscattering from compositions of $N_p = 5000$ point scatterers plotted vs the backscattering angle. Blue and red curves show correlation functions for maximally entangled pure and fully mixed states, respectively. The Schmidt rank is $M = 2$ and the separation angle in incident states is $\Delta\theta = 0.5°$ (a,c) and $\Delta\theta = 8°$ (b,d).

**Appendix III: Angular Speckle Correlation Analysis**

Owing to the wave nature of light, the constructive and destructive interference in the scattering of a classical coherent field from a disordered medium or a rough surface induces a randomly fluctuating scattered intensity in the form of speckle patterns, in the far field or the image plane of the object. Here we calculate the degree of angular speckle correlation, $\gamma_{I_1 I_2}$, by calculating the cross correlation between two speckle patterns [53],

$$\gamma_{I_1 I_2} = \frac{\langle I_1 I_2 \rangle - \langle I_1 \rangle \langle I_2 \rangle}{\sqrt{\left(\langle I_1^2 \rangle - \langle I_1 \rangle^2\right)\left(\langle I_2^2 \rangle - \langle I_2 \rangle^2\right)}}. \tag{11}$$

Using complex Wick's theorem for complex Gaussian statistics this fourth-order moment is equivalent to (see Appendix IV for further details)

$$\gamma_{I_1 I_2} = \frac{\left|\langle A_1 A_2^* \rangle\right|^2}{\langle A_1 A_1^* \rangle \langle A_2 A_2^* \rangle}, \tag{12}$$

in terms of complex backscattered fields, $A_i$, $i \in \{1,2\}$. The two speckle intensities (or fields), $I_1 \left(or\ A_1\right)$ and $I_2 \left(or\ A_2\right)$, are numerically calculated by illuminating the disordered configuration with different incident angles.

Figure. 12 shows the angular dependence of the speckle correlation, numerically calculated using Eqs. (11) and (12) by setting the angle of incidence to $\theta_{inc} = 140°$ for $I_1$ and calculating $I_2$ for incident angles around $140°$. The input fields are tilted plane waves and the medium consists of 5000 point-like scatterers, randomly located inside a cube with edge size $L = 20\lambda$. The blue curve depicts the numerically calculated correlation (Eq. (11)) and the red curve is the simplified calculation using Wick's theorem (Eq. (12)). As expected, a quick decorrelation in speckle is obtained, consistent with experimental observations, commonly used in evaluations of surface roughness [54], [55]. One may note that although the two curves are quite comparable, there are some discrepancies at larger incident angles. Here, the speckle correlation angular width, defined as the full width of the correlation at half-maximum height, of $\Delta\theta_{corr} = 2.25°$ is observed. This angular width is consistent with the extent of the separation angles leading to increased and slightly non-uniform ensemble-averaged angular correlations in the far field (see Figs. 5 and 7 and the discussions in the main text).

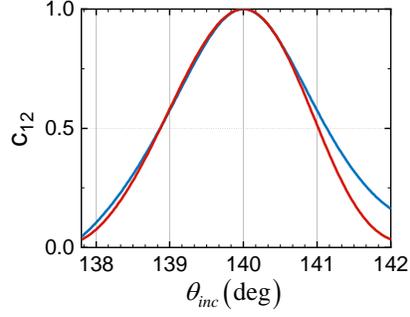

FIG. 12. Angular speckle correlation function for illumination of $N_p = 5000$ scatterers randomly positioned in a cube with edge size $L = 20\lambda$ for incident angles around $140°$. The blue curve corresponds to exact correlation (Eq. (11)) and the red curve shows the correlation using Wick's theorem (Eq. (12)).

**Appendix IV: Analytical Estimations for Complex Gaussian Statistics; Quantum Anticorrelation and Photon Bunching**

Classical circular Gaussian statistics are commonly considered for statistical behavior of scattered complex field amplitudes, i.e., scattering matrix elements, from disordered geometries. This applies assuming adequate statistically independent scattering events. In that case and from random-matrix theory [56], these complex amplitudes may statistically be modeled as independent random variables with identical Gaussian distributions. According to the complex Wick's theorem on the moments of a complex zero-mean multi-variant Gaussian distribution with variables $z_n$ for $(n = 1, 2, ... N)$, the general first moment can be found as [57],

$$E\{\bar{z}_{m_1} \bar{z}_{m_2}, \cdots \bar{z}_{m_s} z_{n_1} z_{n_2} \cdots z_{n_t}\} = \begin{cases} 0 & s \neq t \\ \sum_\pi \left(E\, \bar{z}_{m_{\pi(2)}} z_{n_1}\right)\left(E\, \bar{z}_{m_{\pi(2)}} z_{n_2}\right)\cdots\left(E\, \bar{z}_{m_{\pi(t)}} z_{n_t}\right) & s = t \end{cases}, \quad (13)$$

in which $\pi$ marks a permutation of all the sub-subscripts $\{1, 2, ..., t\}$. Subsequently, and by considering the first and second moments for complex Gaussian scattering elements to be $\langle S_{mn} \rangle = 0$ and $\langle |S_{mn}|^2 \rangle = \sigma^2$, respectively, all the other moments may be determined. The first moments for backscattered single- and two-photon currents are therefore identical for studied cases of maximally entangled pure and fully mixed states, and are equal to

$$\langle I_1(\mathbf{k}_o) \rangle = 2\sigma^2, \quad \langle I_2(\mathbf{k}_o, \mathbf{k}_o) \rangle = 4\sigma^4. \quad (14)$$

The ensemble-averaged angular quantum correlation in backscattering may as well be estimated analytically by substituting the ensemble averages in Eq. (5) with statistical averaging (i.e., mean value) for the complex Gaussian distributions as

$$\overline{C_e(\mathbf{k}_o,\mathbf{k}_o)}\Big|_{Gaussian} = \overline{C_m(\mathbf{k}_o,\mathbf{k}_o)}\Big|_{Gaussian} = \frac{2M}{1+2M}. \tag{15}$$

Provided these assumptions are valid, the quantum correlations depend on the Schmidt rank $M$ and increase by increasing $M$, yet are always less than unity; and hence anticorrelation can clearly be observed. The observed anticorrelation is a manifestation of quantum interference between different paths in a multiple-scattering configuration and does not appear for input coherent states. It is noteworthy that this effect occurs for two-photon input states with both quantum correlations (entanglement) and classical correlations.

At the same time, for two distinct observation modes, the first moment for the two-photon current is $2\sigma^4$ and therefore spatial photon bunching, i.e., $\langle I_2(\mathbf{k}_o,\mathbf{k}'_o)\rangle < \langle I_2(\mathbf{k}_o,\mathbf{k}_o)\rangle$, can be observed. Additionally, for two separate angular modes, by using Wick's theorem, the averaged quantum correlation function for both entangled and mixed states is equal to one half, regardless of the value of $M$. Therefore, the two distinct angular modes are always anticorrelated. Also, the degree of averaged angular correlation for coinciding modes (Eq. (15)) is always greater than the correlation for two separate observation modes. For $M=1$ this 2-channel spatial anticorrelations has been previously observed in quasi-ballistic and very weak disorder regimes [10] by calculating short- and long-range correlation functions and employing random-matrix theory. Furthermore, increasing the degree of disorder and transitioning from weakly disordered regime to Anderson localization is shown to result in an increase followed by a saturation of the 2-channel correlation function.

In addition, for the two photons incident in the same input mode, i.e., single mode Fock state $|2_\mathbf{q}\rangle = 1/\sqrt{2}\, a_\mathbf{q}^\dagger a_\mathbf{q}^\dagger |0\rangle$ as excitation, the single and two-photon currents in coinciding output modes are $2|S_{\mathbf{k}_o,\mathbf{q}}|^2$ and $2|S_{\mathbf{k}_o,\mathbf{q}}|^4$, respectively, resulting in an angular quantum correlation of $\overline{C(\mathbf{k}_o,\mathbf{k}_o)} = 0.5$ regardless of the composition or degree of disorder in the geometry, illustrating an anticorrelation for all geometries.